\documentstyle[epsfig]{mn}

\newif\ifAMStwofonts
\ifoldfss
  \newcommand{\rmn}[1] {{\rm #1}}

  \ifCUPmtlplainloaded \else
    \NewTextAlphabet{textbfit} {cmbxti10} {}
    \NewTextAlphabet{textbfss} {cmssbx10} {}
    \NewMathAlphabet{mathbfit} {cmbxti10} {} % for math mode
    \NewMathAlphabet{mathbfss} {cmssbx10} {} %  "   "    "
  \fi
  \ifAMStwofonts
    \ifCUPmtlplainloaded \else
      \NewSymbolFont{upmath} {eurm10}
      \NewSymbolFont{AMSa} {msam10}
      \NewMathSymbol{\upi}     {0}{upmath}{19}
      \NewMathSymbol{\umu}     {0}{upmath}{16}
      \NewMathSymbol{\upartial}{0}{upmath}{40}
      \NewMathSymbol{\leqslant}{3}{AMSa}{36}
      \NewMathSymbol{\geqslant}{3}{AMSa}{3E}

      \let\leq=\leqslant 
      \let\geq=\geqslant 
    \fi
  \fi
\fi % End of OFSS

\ifnfssone
  \newmathalphabet{\mathit}
  \addtoversion{normal}{\mathit}{cmr}{m}{it}
  \addtoversion{bold}{\mathit}{cmr}{bx}{it}
  \newcommand{\rmn}[1] {\mathrm{#1}}

  \newmathalphabet{\mathbfit} % math mode version of \textbfit{..}
  \addtoversion{normal}{\mathbfit}{cmr}{bx}{it}
  \addtoversion{bold}{\mathbfit}{cmr}{bx}{it}
  \newmathalphabet{\mathbfss} % math mode version of \textbfss{..}
  \addtoversion{normal}{\mathbfss}{cmss}{bx}{n}
  \addtoversion{bold}{\mathbfss}{cmss}{bx}{n}
  \ifAMStwofonts
    \ifCUPmtlplainloaded \else
      \UseAMStwoboldmath
      \makeatletter
      \new@mathgroup\upmath@group
      \define@mathgroup\mv@normal\upmath@group{eur}{m}{n}
      \define@mathgroup\mv@bold\upmath@group{eur}{b}{n}
      \edef\UPM{\hexnumber\upmath@group}
      \new@mathgroup\amsa@group
      \define@mathgroup\mv@normal\amsa@group{msa}{m}{n}
      \define@mathgroup\mv@bold\amsa@group{msa}{m}{n}
      \edef\AMSa{\hexnumber\amsa@group}
      \makeatother
      \mathchardef\upi="0\UPM19
      \mathchardef\umu="0\UPM16
      \mathchardef\upartial="0\UPM40
      \mathchardef\leqslant="3\AMSa36
      \mathchardef\geqslant="3\AMSa3E

      \let\leq=\leqslant 
      \let\geq=\geqslant 
    \fi
  \fi
\fi % End of NFSS release 1

\ifnfsstwo
  \newcommand{\rmn}[1] {\mathrm{#1}}

  \DeclareMathAlphabet{\mathbfit}{OT1}{cmr}{bx}{it}
  \SetMathAlphabet\mathbfit{bold}{OT1}{cmr}{bx}{it}
  \DeclareMathAlphabet{\mathbfss}{OT1}{cmss}{bx}{n}
  \SetMathAlphabet\mathbfss{bold}{OT1}{cmss}{bx}{n}
  \ifAMStwofonts
    \ifCUPmtlplainloaded \else
      \DeclareSymbolFont{UPM}{U}{eur}{m}{n}
      \SetSymbolFont{UPM}{bold}{U}{eur}{b}{n}
      \DeclareSymbolFont{AMSa}{U}{msa}{m}{n}
      \DeclareMathSymbol{\upi}{0}{UPM}{"19}
      \DeclareMathSymbol{\umu}{0}{UPM}{"16}
      \DeclareMathSymbol{\upartial}{0}{UPM}{"40}
      \DeclareMathSymbol{\leqslant}{3}{AMSa}{"36}
      \DeclareMathSymbol{\geqslant}{3}{AMSa}{"3E}

      \let\leq=\leqslant 
      \let\geq=\geqslant 
    \fi
  \fi
\fi % End of NFSS release 2

\ifCUPmtlplainloaded \else
  \ifAMStwofonts \else % If no AMS fonts
    \def\upi{\pi}
    \def\umu{\mu}
    \def\upartial{\partial}
  \fi
\fi

\title{Lithium in LMC Carbon Stars}
\author[D.Hatzidimitriou et al.]
       { D.~Hatzidimitriou,$^1$ D.H.~Morgan,$^2$ R.D.~Cannon,$^3$
        B.F.W.~Croke$^4$ \\
       $^1$ Department of Physics, University of Crete, Heraklion, Greece \\
       $^2$ Institute for Astronomy, University of Edinburgh, Royal
            Observatory, Blackford Hill, Edinburgh  EH9 3HJ, UK \\
       $^3$ Anglo-Australian Observatory, PO Box 296, Epping, NSW 2121,
            Australia  \\
       $^4$ Integrated Catchment Assessment and Management Centre,
            Centre for Resource and Environment Studies, \\
            Australian National University, Canberra, ACT 0200, Australia}
\date{Accepted
      Received
      in original form }

\pagerange{\pageref{firstpage}--\pageref{lastpage}}
\pubyear{2000}

\begin{document}

\maketitle

\label{firstpage}

\begin{abstract}
Nineteen carbon stars that show lithium enrichment in their
atmospheres have been discovered among a sample of 674 carbon
stars in the Large Magellanic Cloud. Six of the Li-rich carbon
stars are of J-type, i.e., with strong $^{13}$C isotopic features. 
No super-Li-rich carbon stars were found. The incidence of lithium 
enrichment among carbon stars in the LMC is much rarer than in 
the Galaxy, and about five times more frequent among J-type than 
among N-type carbon stars. The bolometric magnitudes of the Li-rich 
carbon stars range between $-$3.3 and $-$5.7. Existing 
models of Li-enrichment via the hot bottom burning process 
fail to account for all of the observed properties of the
Li-enriched stars studied here.

\end{abstract}


\section{Introduction}
%\setcounter{paragraph}{0}
%\paragraph{}
Lithium is a sensitive indicator of nucleosynthesis and
convective dredge-up in stars, as it is easily produced as well as easily
destroyed in stellar interiors.

%\paragraph{}
It has been known for a number of decades (e.g. Torres-Peimbert \&
Wallerstein, 1966) that some Galactic asymptotic giant branch
(AGB) stars possess large surface lithium abundances,
reaching values one or two orders of magnitude higher than the
cosmic value. More recently, Li-rich AGB stars have also been
identified in other Local Group galaxies, namely the Large and Small
Magellanic Clouds (LMC and SMC) and M31.
%(Smith \& Lambert
%\shortcite{smith89}; Smith \& Lambert \shortcite{smith90}).
The great majority of Li-rich AGB stars discovered to date in the
Magellanic Clouds have oxygen-rich rather than carbon-rich
atmospheres \cite{smith95} (hereafter SPLL95), although the opposite 
may be true in the Galaxy (Abia et al., 1993, Abia \& Isern 1997).

%\paragraph{}
The lithium enrichment of the atmospheres of {\em oxygen-rich} AGB
stars (i.e., stars with C/O\,$<$\,1) is relatively well understood
within the framework of the hot bottom burning (HBB) process
during which the bottom of the convective envelope of the
star reaches the top layers of the hydrogen-burning shell.
Sackmann \& Boothroyd \shortcite{sackmann92} showed that in the
intermediate mass range (4--6M$_{\odot}$) the Cameron-Fowler $^7$Be
transport mechanism \cite{cameron71} is very
effective at producing $^7$Li (via the decay of $^7$Be) and can
account for the super-Li-rich, oxygen-rich AGB stars found by
Smith \& Lambert \shortcite{smith90} in the Magellanic Clouds.

%\paragraph{}
{\em Carbon-rich} AGB stars (C/O$>\,$\,1), (i.e., {\em carbon stars}),
also display enhanced lithium abundances. Very
large equivalent widths (W$_{6707}\sim10$\AA) of
the Li\,{\sc i}\,$\lambda$6707 resonance doublet have been found in a few
Galactic carbon stars: examples are WZ~Cas and WX~Cyg which are
classified as \mbox{N7 (C$_2$ 2)} and \mbox{J6 (C$_2$ 3$^-$)} by Keenan
\shortcite{keenan93} and Barnbaum, Stone \& Keenan
\shortcite{barnbaum96} respectively. However, most Galactic carbon
stars have much weaker lithium lines with W$_{6707}<0.5$\AA \/ and some have
0.5\AA\,$<$\,W$_{6707}<1$\AA \/ \cite{torres66}.
In the Magellanic Clouds, on the other hand, carbon stars that display
lithium-enriched atmospheres seem to be rarer and generally less extreme
in terms of lithium abundance, compared with their Galactic counterparts
or with the oxygen-rich AGB stars within the Clouds
(e.g., SPLL95) mentioned in the previous paragraph.
In the most comprehensive study yet of Li-rich AGB stars in the Magellanic
Clouds by SPLL95, only four Li-rich carbon 
stars were recorded in the LMC and just two in the SMC.

%\paragraph{}
The study of Li-rich carbon stars present difficulties both
observationally and theoretically.  On the observational side,
severe measurement difficulties arise from the blending of the
lithium lines with the complex array of lines in the red 
A\,$^2\Pi$--X\,$^2\Sigma$ CN system. This heavy line blanketing also 
means that equivalent
width measurements are generally made relative to a
pseudo-continuum rather than the true continuum. From the
theoretical viewpoint, the coexistence of carbon and lithium in
the atmosphere of an AGB star is difficult to achieve, at least in
the hot-bottom-burning scenario, since the conditions required to
ignite lithium production  also lead to fast CNO processing at the base
of the envelope, with the subsequent destruction of the $^{12}$C that was
previously conveyed to the stellar surface.  Ventura, D'Antona \& Mazzitelli 
\shortcite{ventura99}, however, showed that, for a certain range of stellar
masses corresponding to a very narrow range in bolometric
magnitude of $-5.95<M_{\rm bol}<-5.75$, carbon can survive long enough
in the convective envelope to keep C/O\,$>$\,1, despite the activation
of the Cameron-Fowler mechanism. At the same time, the
$^{13}$C/$^{12}$C ratio increases and the star displays the
spectral characteristics of a J-type carbon star which is also
Li-rich. These models, however, cannot explain the existence of
the fainter (by up to about one magnitude) Li-rich carbon stars
found by SPLL95 in the Magellanic Clouds, as
well as by Abia et al. \shortcite{abia93} in the Galaxy.  Abia \& Isern
\shortcite{abia97} proposed an alternative mechanism to HBB, involving deep
mixing, operating in low-mass carbon stars (M\,$<$\,2M$_{\odot}$) which
can account for the fainter Li-rich and $^{13}$C-rich (J-type)
carbon stars, found in the Galaxy and in the Magellanic Clouds.  A more
detailed description of the various models will be discussed later
in the light of the new observations described in the present
study.

%\paragraph{}
For Li-rich carbon stars, the derivation of reliable lithium abundances,
the acquisition of adequate statistics of their occurrence, as
well as an understanding of their exact evolutionary stage, are
all outstanding and open issues which are important not only for
late-stage stellar evolution and nucleosynthesis, but also for modelling 
the present-day lithium abundance in our Galaxy and other galaxies
(e.g., Travaglio et al.
\shortcite{travaglio01}, Ventura et al. \shortcite{ventura00},
Romano et al. \shortcite{romano01}). It must be emphasized that
the study of lithium production in external galaxies such as the
Magellanic Clouds is particularly important in this respect, since
lower metallicity and a different evolutionary history would
affect the yield of lithium on galactic scales.

%\paragraph{}
Following the completion of a catalogue of 7760 carbon stars in
the Large Magellanic Cloud by Kontizas et al.
\shortcite{kontizas01} (see also Dapergolas et al.
\shortcite{dapergolas96}), an extensive programme of spectroscopy
of carbon stars has been started \cite{cannon99}
using the 2dF multi-object spectroscopic facility on the Anglo-Australian 
Telescope (AAT) \cite{lewis02}.  The subject of the present paper
is the detection of the Li\,{\sc i}\,$\lambda$6707 resonance doublet 
in the spectra of 19 out of the 674 carbon stars studied.  This is the 
first time that such a large homogeneous spectroscopic sample of 
carbon stars has been studied in the context of lithium enrichment 
in any galaxy.  Having 
homogeneous spectra for such a large sample of carbon stars, all at 
practically the same distance and with little reddening, provides a 
firm basis for making stronger statistical statements about the LMC 
carbon stars than can be done for (current) Galactic, or other 
Magellanic Cloud samples. The paper is organized as follows: Section 2 
describes the observations, Section 3 describes the basic characteristics
of the spectra obtained, while Section 4 outlines the method used
to identify and measure the equivalent widths of the
Li\,{\sc i}\,$\lambda$6707 resonance doublet. In Section 5,
the photometric properties of the identified Li-rich carbon stars
are investigated, while in Section 6, the results are compared with 
those of previous studies, and some theoretical implications are
discussed.

\section{Observations}

%\setcounter{paragraph}{0}
%\paragraph{}
Most of the carbon stars observed were selected from the Kontizas et al.
\shortcite{kontizas01} catalogue of LMC carbon stars: {\sc simbad} 
identifier -- KDM.  This catalogue 
was constructed by identifying the strong (1,\,0) and (0,\,0) C$_2$
Swan bands (at 4737\AA \/ and 5165\AA \/ respectively) in carbon stars 
seen in low-dispersion spectra on an objective-prism survey taken in 
the blue$-$green with the 1.2-m UK Schmidt Telescope. 

%\paragraph{}
The spectroscopic observations were obtained at the AAT on 21 and 22 
January 1998, using the 2dF multi-fibre spectroscopic facility, which 
enables up to 400 spectra to be recorded simultaneously, in a pair of 
identical fibre-fed spectrographs fitted with TEK 1024 CCDs \cite{lewis02} 
-- see http://www.aao.gov.au/2df/.  The target
stars covered four $2^\circ$ diameter fields. The spectra were
centred on the $\Delta\nu$\,=\,+2 Swan bands near 6200\AA \/ and
covered the wavelength range 5675--6785\AA.  Subsequent 2dF observations 
of more Magellanic Cloud carbon stars were taken over a slightly different 
wavelength range which included the strong (0,\,1) Swan band at 5635\AA \/ 
but excluded the Li\,{\sc i}\,$\lambda$6707 line.

%\paragraph{}
The 1200R gratings were used, yielding spectra with 1.1\AA \/ per
pixel or an effective resolution of $\sim2.5$\AA. Several exposures were taken 
for each field, typically 3\,$\times$\,900\,s or 2\,$\times$\,1200\,s,
together with offset sky, arc and flat field exposures.
The data were reduced with the AAO's {\sc 2dfdr} data reduction package 
\cite{bailey01} and the {\sc iraf} package and yielded spectra with 
signal-to-noise 
ratio (S/N) usually around 30 per pixel.  Full details of the observations 
are given by Cannon et al. (in preparation).

%\paragraph{}
Relative radial velocities of all the stars were obtained by using a 
cross-correlation technique in {\sc figaro}.  The high signal-to-noise 
ratio (S/N) usually attained for most stars and the large number of 
spectral features resulted in an internal precision of about 1/20 pixel 
though with external accuracies closer to 1/10 pixel.

%\begin{table}
%\begin{center}
%\caption{Details of the 2dF Obervations}
%\begin{tabular}{cccccc}
%\hline
%Field & Stars & RA & Dec & CCD &  date \\
%      &       &    &     &                    &      \\
%6     & 164   & $5.50^{\rmn h}$ & $-71.4^\circ$ & 1 &  21 Jan 1998 \\
%7     & 112   & $5.33^{\rmn h}$ & $-67.2^\circ$ & 1 &  21 Jan 1998 \\
%8     & 112   & $5.50^{\rmn h}$ & $-73.2^\circ$ & 1 &  21 Jan 1998 \\
%9     & 165   & $5.10^{\rmn h}$ & $-69.8^\circ$ & 1 &  22 Jan 1998 \\
%9     & 166   & $5.10^{\rmn h}$ & $-69.8^\circ$ & 2 &  22 Jan 1998 \\
%\hline
%
%\end{tabular}
%\end{center}
%\end{table}

\section{Spectral Analysis}

\subsection{General considerations}

%\setcounter{paragraph}{0}
%\paragraph{}
The spectra have remarkable similarities due to the abundance of CN lines.
They can be classified, however, in four main groups, which will be used
to aid the analysis. These are: (1) typical N5-type carbon stars, with strong
Swan bands and CN bands,  (2) stars with weak
Swan bands near 6200\AA \/ due to higher temperature and/or lower carbon
abundance, (3) stars with very weak carbon features, (4) J-stars with
strong $^{13}{\rmn C}^{12}{\rmn C}$ and $^{13}{\rmn C}^{14}{\rmn N}$ bands
as well as the normal $^{12}{\rmn C}^{12}{\rmn C}$ and
$^{12}{\rmn C}^{14}{\rmn N}$ bands.  As noted earlier, identification 
of the J-type stars is important in the context of the theories of lithium
enrichment in carbon stars.  Details of the precise method used to 
identify the J~stars are given by Morgan et al. \shortcite{morgan03};
but it is sufficient here to note that the method selects stars with 
a similar level of $^{13}{\rmn C}$ to the standard definition of a J~star,
which is equivalent to $^{12}{\rmn C}^{13}{\rmn C}<15$ \cite{abia97}.
Allocation of the other stars to the first three groups was made by simple
visual inspection.  Examples of the spectral groups can also be found 
in Barnbaum et al. \shortcite{barnbaum96}.
(1: TV\,Lac; 2: BD\,+2$^\circ$3336 and Z\,Psc; 3: HD\,123821
and HD\,76846; 4: HD\,10636 and EU\,And).  Each observed star 
(with sufficient S/N) was assigned to one of 
these four groups.  The numbers of stars in the four groups were 
486, 92, 23 and 66 respectively making a total number of 674 stars.
%The numbers of stars in the four groups were 490, 92, 23 and 69 respectively.
%There were 48 other stars observed but not considered, usually because they
%had insufficient signal due to variability or coordinates affected by close
%companions.The number of used spectra was 674.
%THE TOTAL NUMBER OF STARS OBSERVED IS THEREFORE 674+48=722, WHILE THE TOTAL
%NUMBER OF STARS APPEARING IN TABLE 1 IS 719. PLEASE CHECK!

%\paragraph{}
The spectral range used in the following analysis was limited to
6690--6730\AA. Each spectrum was corrected for its radial velocity
using a grid of interpolated wavelength points to allow this to be
done to the sub-pixel accuracy of the velocities.  It was then
normalized using mean counts in bands at 6695--6703\AA \/ and
6712--6720\AA.

%\paragraph{}
A mean spectrum was constructed for each one of the four spectral groups.
The mean spectra of N and J~stars (Groups 1 and 4) are shown in
Figs~1a\,\&\,1d respectively.  The mean spectra of the carbon-weak stars
(Groups 2 and 3) are like that for Group 1 but with weaker features.
The J-star combined spectrum is clearly different
from the N-star spectrum because of the presence of $^{13}$CN bands.
The spectrum of the LMC carbon star BMB~R-46 taken from SPLL95 is shown in 
Fig.~1b; the similarity between it and the N-star spectrum of Fig.~1a
is striking.  The Galactic carbon star SY\,Eri, which is
classified by Barnbaum et al. \shortcite{barnbaum96} as 
\mbox{N5 (C$_2$\,5)}, is shown at a higher resolution in Fig.~1c \cite{abia93}
and is again similar to the combined N-star spectrum.
However, there are differences: Ca\,{\sc i}\,$\lambda$6717 is strong in 
SY\,Eri but is not apparent in either of the LMC stars and there are more
features in the region 6702--6710\AA \/ of the Galactic star.  These
differences are due in part to the lower resolution of the LMC spectra 
and also to the lower metallicity of the LMC.  Indeed, the strength of 
the calcium line in SY\,Eri is consistent with solar abundance \cite{abia93}.

%\paragraph{}
%For demonstration purposes, Fig.~1d shows a low-resolution synthetic
%CN spectrum for two temperatures -- 3000\,K and 5000\,K, and 
%a high-resolution one for 3000\,K.  These synthetic spectra are simply 
%the total absorption coefficient computed by adding the individual 
%CN lines broadened by a gaussian of {\sc fwhm} 2.5\AA \/ (or 1.0\AA), 
%to match the resolution of the 2dF spectra.  The energy levels were 
%calculated following Fay, Marenin \& van Citters \shortcite{fay71}, 
%using their equations (1)\,--\,(7) and
%tabulated molecular constants.  Since complexities such as perturbations 
%were ignored during these calculations, the resulting wavelengths of 
%transitions between the levels differed slightly from the measured 
%values of Davis \& Phillips \shortcite{davis63}.  A correction (quadratic 
%fit) was applied to the calculated values to match the measured ones.  
%Intensities were calculated following Marenin \& Greene
%\shortcite{marenin72}, assuming a Boltzmann distribution, Holn-London
%factors from Schadee \shortcite{schadee64}, relative band oscillator
%strengths from Lambert \shortcite{lambert68} and total internal partition
%functions from Tatum \shortcite{tatum66}.
%Although the synthetic spectra were not constructed with the purpose of
%carrying out a quantitative comparison with the observed spectra, there
%appears to be a fairly good qualitative agreement between modelled
%and observed (Group 1) spectral features.

\begin{figure}
\vspace{15.0cm}
\includegraphics{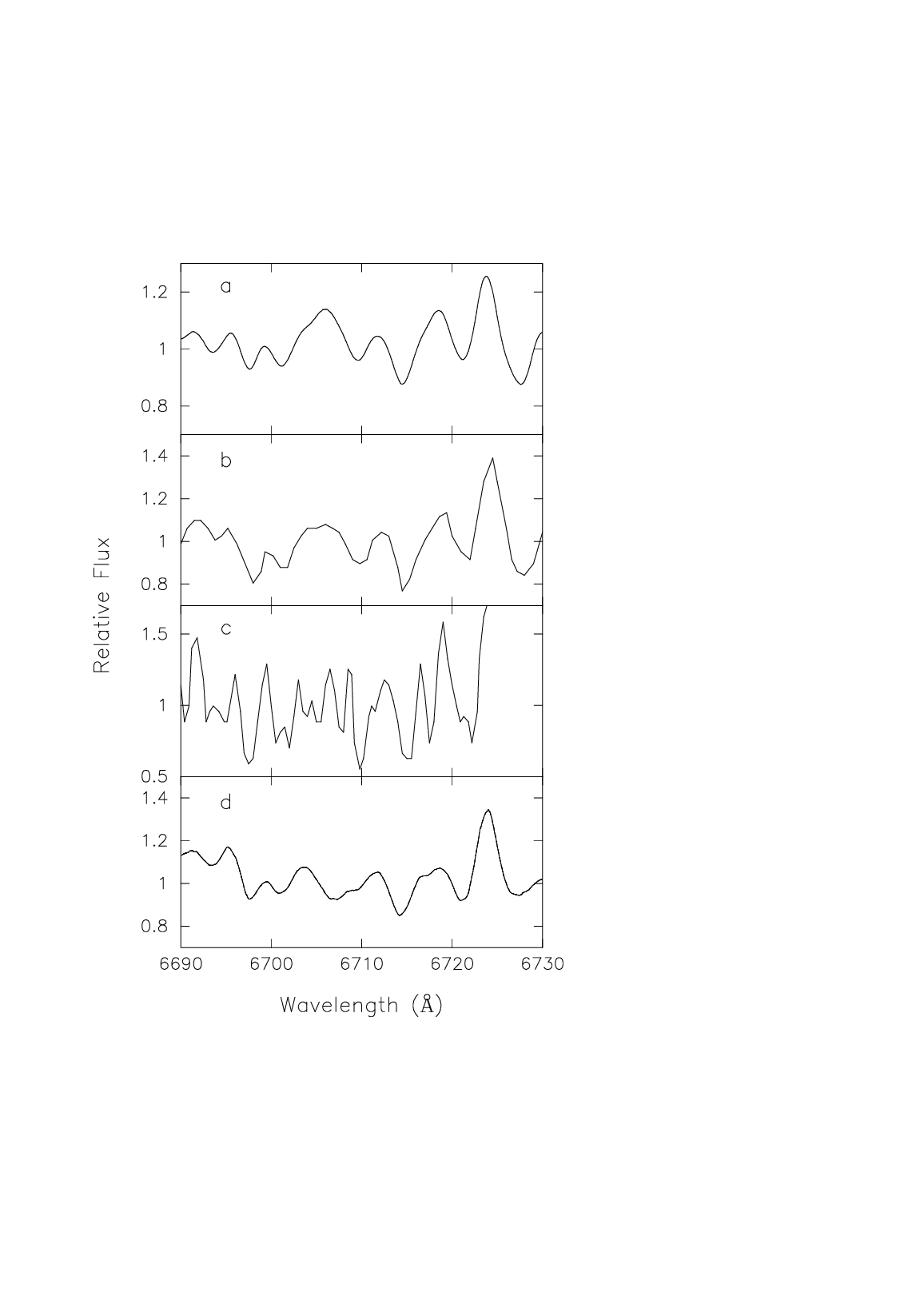}
\caption{Spectra in the region around Li\,{\sc i}\,$\lambda$6707.
(a) mean spectrum of Group 1 ($\sim$N5) stars;
(b) LMC star BMB R-46 from SPLL95;
(c) Galactic star SY\,Eri (type N5(C$_2$\,5)) from Abia et al. (1993);
%(d). synthetic CN spectrum for \mbox{T$_{\rm eff}$ = 3000\,K} (continuous and 
%dashed (higher resolution) lines) and \mbox{T$_{\rm eff}$ = 5000\,K} 
%(dotted line), 
(d) mean spectrum of Group 4 (J) stars.}
\label{spec}
\end{figure}

\subsection{Identification of Li\,{\sc i}\,$\lambda$6707}

%\setcounter{paragraph}{0}
%\paragraph{}
The Li\,{\sc i} resonance doublet at 6707.8\AA \/ lies in a spectral 
region which is heavily affected by many atomic and molecular features.
These are mainly CN lines, but molecular 
lines of C$_2$ and TiO are also present near the Li doublet.  However, 
the C$_2$ lines are of minor importance compared with the CN lines 
since at all optical depths the CN concentration is a factor of 7
to 100 times higher than the C$_2$ concentration \cite{abia93}.  Moreover,
since C/O\,$>$\,1, CO formation is favoured against TiO.
The presence of strong CN bands makes it difficult to use simple visual
means to select stars with enhanced lithium, and even harder to measure the
strength of the line.  

\begin{figure}
\vspace{16.0cm}
\includegraphics{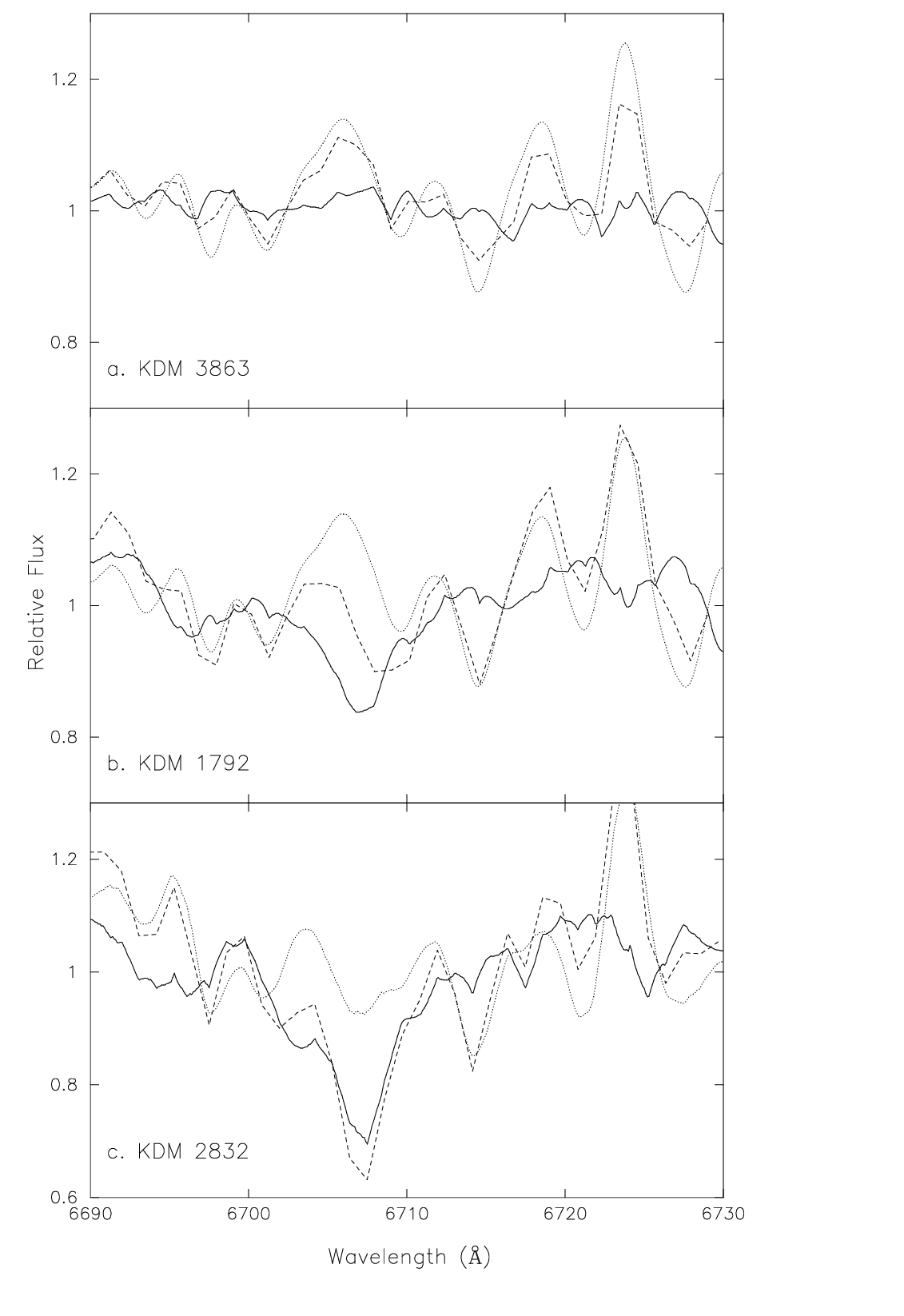}
\caption{Normalized spectra of three carbon stars.  The dashed line is 
the spectrum of a star, the dotted line is the mean spectrum of all stars
in the same spectral group and the solid line is the difference between 
the spectrum and the scaled mean spectrum.  (a). the N~star (Group~1) 
KDM\,3863 which has no lithium line; (b). the N~star 
KDM\,1792 which does show a lithium line; (c). the J~star 
(Group~4) KDM\,2832 which has a strong lithium line.  The 
scaling factors applied to the mean spectra were 0.6, 1.0 and 0.9
respectively.}
\label{montage}
\end{figure}

The method chosen here to identify the lithium line
in a particular star was to subtract the mean observed 
spectrum (described above) of the spectral group appropriate 
for the star in question, scaled 
according to the strength of the CN bands in the individual spectrum.  
The scaling factor was determined by minimizing the difference 
between the spectrum and the scaled mean.  The subtracted spectrum
is mainly that of CN, but includes contributions from metal lines and 
other molecular bands.  The subtraction process is designed to reveal 
only those features that deviate significantly from the scaled mean 
spectrum. Fig.~\ref{montage} shows examples of this process for three 
stars: two typical N~stars (Group 1), one with 
and the other without a lithium line, and a J~star (Group 4) with 
a strong lithium line.  The relative smoothness of the residual spectra 
beyond the bounds of the lithium line and seen throughout the entire 
upper panel shows that this is a valid procedure which takes account 
of the strong star-to-star variations in the overall strengths of the
CN bands, presumably depending on both temperature and carbon and 
nitrogen abundances.

%\paragraph{}
Residual spectra such as those shown as solid lines in Fig.~\ref{montage}
were constructed for every observed star.  
Lines in the residual spectra were then sought as follows: a
second-order polynomial was fitted to the residual spectrum using the
least-squares method in order to
define the `continuum' and all maxima and minima (potential
emission and absorption lines respectively) about this fit were
identified. A gaussian, superimposed on the quadratic fit, was
then fitted to each maximum and minimum allowing its central
wavelength ($\lambda_{cen}$), height and full width at half maximum 
({\sc fwhm}) to vary. The equivalent widths of the
identified (both emission and absorption) lines (W$_{\lambda}$)
were then measured. It should be emphasized that these
equivalent widths should be used in a relative sense only, due to
their strong dependence on the way the continuum is defined. Other
authors use a pseudo-continuum, so equivalent widths from different 
sources are not directly comparable. 

\begin{figure}
\vspace{9.0cm}
\includegraphics{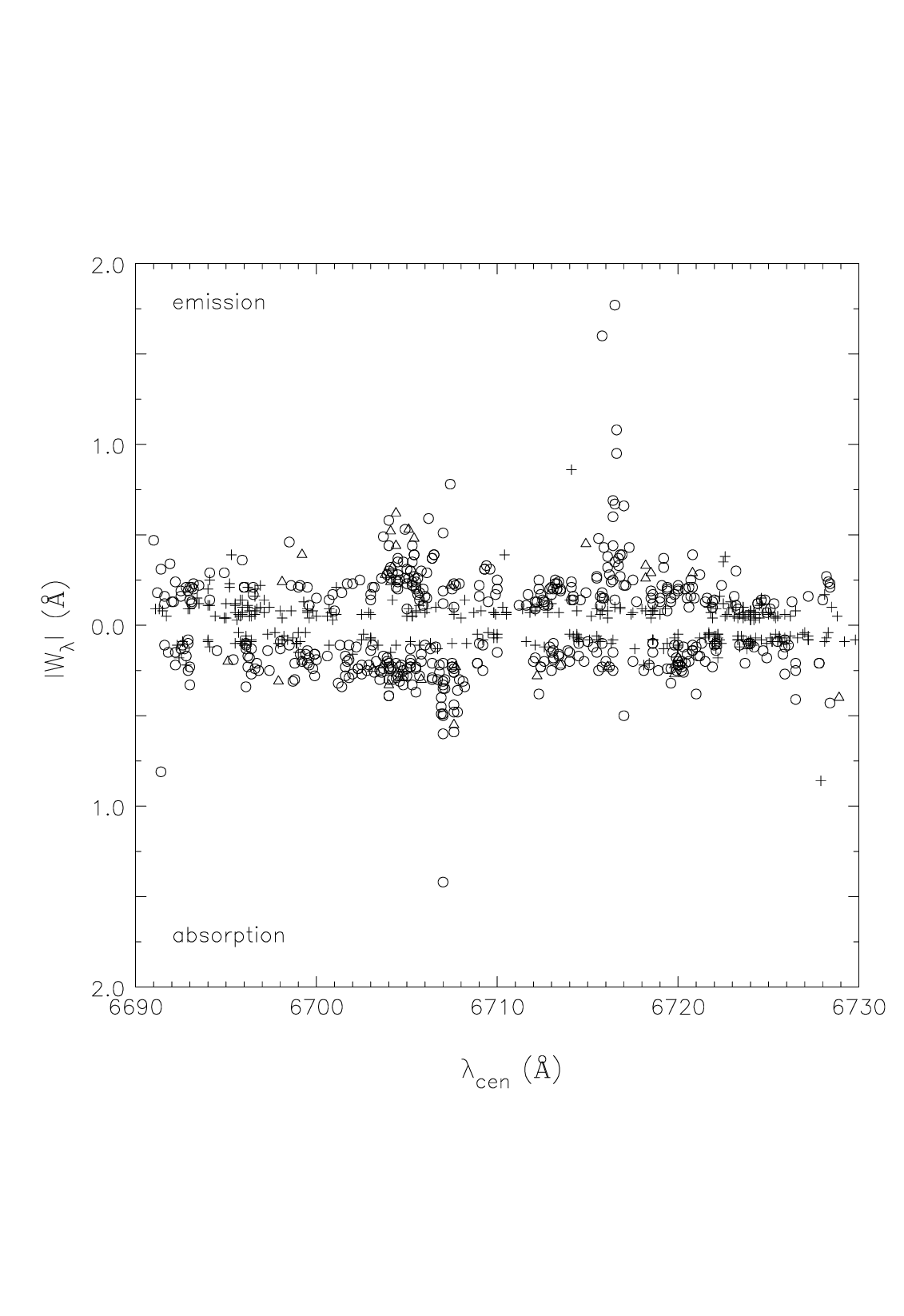}
\caption{Equivalent widths (W$_{\lambda}$) of features detected in the 
residual spectra.  Absorption features appear in the lower half of
the figure; emission features appear in the upper half.  All W$_{\lambda}$ 
are shown as positive values, the ordinate running positively from zero 
in each half of the diagram.  The symbols denote the {\sc fwhm} of the 
fitted gaussian line: {\sc fwhm}\,$<$\,2.0\AA \/ (plus), 
2.0\AA\,$\leq$\,{\sc fwhm}\,$\leq$\,5.0\AA (circle) and 
{\sc fwhm}\,$>$\,5.0\AA \/ (triangle).  For clarity, only one in three 
points is plotted for $|W_\lambda|<0.2$\,\AA.}
\label{lines}
\end{figure}

%\paragraph{}
Clearly, a spectrum with poor signal-to-noise ratio will generate random 
`lines'.
It turns out that the number of detected lines starts to rise significantly
once the count level falls below $\sim$500 counts per pixel.  Consequently, 
53 stars with fewer than 500 counts per pixel were excluded from further 
consideration.

The derived W$_{\lambda}$ are plotted against 
$\lambda_{cen}$ in Fig.~\ref{lines}.  What appears as a gap 
around $W_{\lambda}=0$ is merely a consequence of the limit applied 
to the line search procedure.  Absorption features appear in the lower
part of the figure; emission features appear in the upper half. 
The fairly uniform
distribution of points with $|W_{\lambda}| \leq$~0.25\AA \/ demonstrates
the noise level of the identification procedure, which results
from the S/N in the original spectra, the spectral
mismatch between star and template and/or velocity errors. Most
of the features in this region of the diagram have 
{\sc fwhm}\,$\simeq$\,2\AA \/ and are indeed unlikely to correspond to true
absorption (or emission) features given the spectral
resolution of 2.5\AA.  However, there are a few outlying points,
both below (absorption) and above (emission) this band, which have
$|W_{\lambda}| \geq$\,0.3\AA \/ and which were selected for further
investigation.  In particular, there is a peak in the distribution of 
absorption features at 6708\AA \/ and emission peaks at 6717\AA \/
and 6706\AA.

%\paragraph{}
The emission features at 6717\AA \/ are due to 
[S\,{\sc ii}]\,$\lambda$6717, most probably originating in 
line-of-sight nebulosity in the LMC.  This supposition is corroborated by 
the fact that the same stars also show strong H$\alpha$ 
emission, as one would expect.  The occasional appearance of nebular 
emission lines in the 2dF data is not surprising because there is 
patchy nebulosity across the LMC field, but the sky subtraction was done 
using the mean signal from a relatively small number of sky fibres chosen 
to lie in clear areas. 
However, the [S\,{\sc ii}]\,$\lambda$6717 emission lines do 
help place limits on the acceptable values of the gaussian {\sc fwhm}.  
Most of these have 
2\AA\,$<$\,{\sc fwhm}\,$<$\,3\AA, and none has {\sc fwhm}\,$<$\,2\AA \/
or {\sc fwhm}\,$>$\,4.5\AA.  Consequently, all detections with 
{\sc fwhm}\,$<$\,2\AA \/ were excluded from further consideration.  
In practice, no features rejected in this way were close to the 
position of the lithium line.  Since small spectral mismatches in the
region around the lithium line can affect the wings of the detected
lithium feature, as is the case with the strong line shown in
Fig.~\ref{montage}c, an upper limit for the line width was set 
at {\sc fwhm}\,$\leq$\,5\AA.

\begin{figure}
\vspace{10.0cm}
\includegraphics{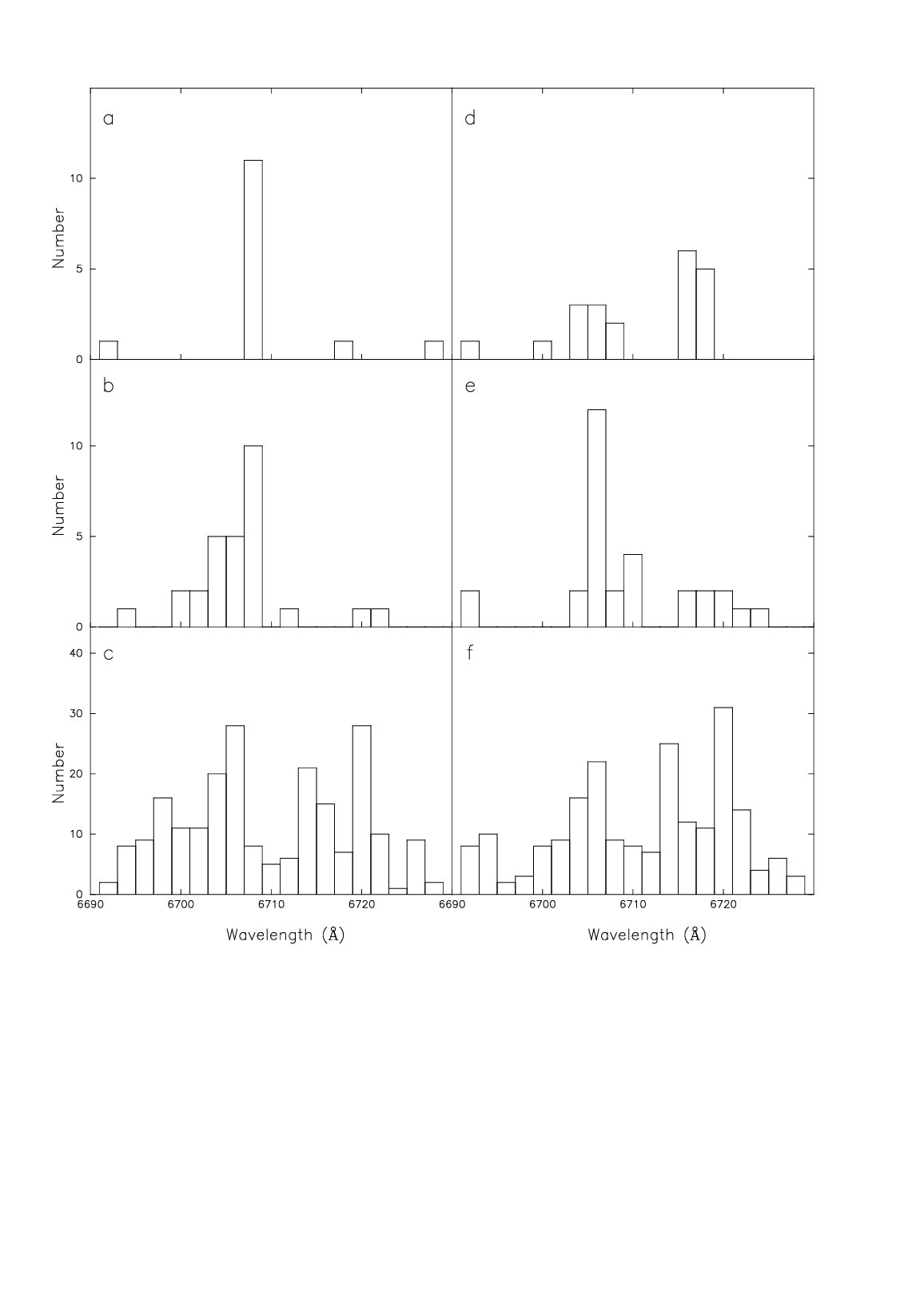}
\caption{Histograms of `lines' detected in the residual spectra.  Panels
(a), (b) and (c) are for absorption `lines' with W$_\lambda$\,$>$\,0.4\AA,
0.3\,$<$\,W$_\lambda$\,$<$\,0.4\AA \/ and W$_\lambda$\,$<$\,0.3\AA \/
respectively;  Panels (d), (e) and (f) are for emission `lines' in the same
three ranges of W$_\lambda$.}
\label{hist}
\end{figure}

%\paragraph{}
Another way to study these results is to consider Fig.~\ref{hist} which 
shows histograms of the identifications plotted in Fig.~\ref{lines} which 
lie in the range 2\AA\,$\leq$\,{\sc fwhm}\,$\leq$\,5\AA.  The 40\AA \/ 
wavelength range is divided into 2\AA-wide bins.  The three left-hand 
panels are for absorption features and 
the three right-hand panels are for emission features.  The individual 
histograms are for identifications with W$_\lambda$\,$>$\,0.4\AA,
0.3\,$<$\,W$_\lambda$\,$<$\,0.4\AA \/ and W$_\lambda$\,$<$\,0.3\AA. 
The predominance of the peak at 6708\AA \/ in the stronger absorption 
features is obvious (Fig.~\ref{hist}a):  it contains eleven stars.
one of which, KDM\,2832, has a very strong absorption line at 6707\AA \/ and 
stands out clearly from the rest in Fig.~\ref{lines}.  Its spectrum 
is plotted in Fig.~\ref{montage}c and shows a Li\,{\sc i}\,$\lambda$6707 
that is strong enough to be seen even without the 
subtraction of the template spectrum.   Of the remaining three features
in Fig.~\ref{hist}a, one appears at 6691\AA \/ in star KDM\,2379, 
another is the blend of two smaller features at 6717\AA \/ in star 
KDM\,3389 and the third is the consequence of a bad pixel.  Since none 
of the bins in Fig.~\ref{hist}a apart from the 6708\AA \/ bin has more than 
one star, it is reasonable to assume that no more than one of the eleven 
stars in the 6708\AA \/ bin is there by chance.  With only two out of 
18 bins having single stars, the likelihood that this is happening is only
$\sim$0.11.

The same peak can be seen in the weaker lines of
Fig.~\ref{hist}b, but here there are also significant numbers of
lines in the two neighbouring bins at lower wavelengths.  Nevertheless,
the histogram certainly reflects a non-random distribution.  It is 
unlikely that the detections shortward of 6708\AA \/ are due to
enhancements of individual metal lines because related metal lines are 
found in other wavelength bins as
well \cite{barnbaum94} and are relatively weak in LMC stars.   It is 
more likely that they are associated with a general spectrum 
characteristic which also produces the relatively high numbers of weak 
lines found (Fig.~\ref{hist}c) at the same two bins and also gives rise to
a number of emission features at 6706\AA \/ (Fig.~\ref{hist}e).  This
could be due to differences in the relative strengths of the 
$^{12}{\rmn C}^{12}{\rmn C}$, $^{13}{\rmn C}^{12}{\rmn C}$,
$^{12}{\rmn C}^{14}{\rmn N}$ and $^{13}{\rmn C}^{14}{\rmn N}$
molecular bands between some stars and the mean spectrum.  In
contrast, very few very weak lines are found at 6708\AA.  Nor are many
emission lines found at 6708\AA.  If, as is likely, the weak
lines give a measure of how errors in the method generate lines,
two of the ten lines in the 6708\AA \/ bin in Fig.~\ref{hist}b are 
expected to have been caused in this way.  If they do not, then the 
number of false detections at 6708\AA \/ 
(at 0.3\,$<$\,W$_{\lambda}$\,$<$\,0.4\AA) could be as high as the number 
seen in the neighbouring bin at 6706\AA, i.e., five out of the ten
detections at 6708\AA \/ could be false.

\begin{figure}
\vspace{6.5cm} 
\includegraphics{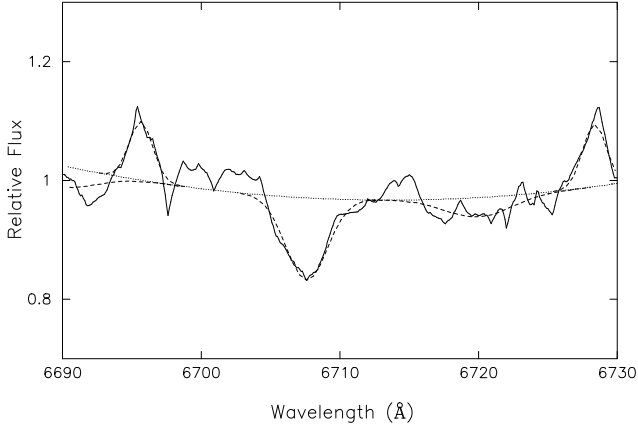} 
\caption{Normalized spectrum of an J-type
carbon star (KDM\,2155) with a lithium line. The superimposed 
dotted and dashed lines are the second-oreder `continuum' and the fitted 
gaussians respectively.}
\label{Nstar_with_li}
\end{figure}

%\paragraph{}
Fig.~\ref{Nstar_with_li} is an example of a J~star with a modest 
detection of Li\,{\sc i}\,$\lambda$6707 with W$_{6707}$\,=\,0.48\AA.
(This can be compared with the strongest lithium line detected in our 
sample, shown in Fig.~\ref{montage}c.) The
superimposed lines show the quadratic `continuum' and the fitted
gaussians.   This particular example includes gaussians fitted to
two broad shallow absorption features and two narrow emission
features; all these have {\sc fwhm} values that are inappropriate for
real lines.  Although very clear in this figure, the line at 6707\AA \/ 
cannot be seen directly in the original spectrum, but is manifested as a 
lowering of the maximum that usually arises between the groups of 
CN lines at 6702\AA \/ and 6709\AA \/ (see Fig.~\ref{spec}). Changing 
the `continuum' fit from quadratic to cubic or linear changes the set
of selected lines slightly and generally reduces (cubic) or
increases (linear) the equivalent widths. However, most lines
selected as lithium are recovered in each case.

%\paragraph{}
At this point it is important to examine other possible sources of the
features identified here. The CN bands coincident with 
Li\,{\sc i}\,$\lambda$6707 are: Q$_1$(28)+Q$_2$(29) of 
the (12,\,7) band, Q$_2$(44) of the (6,\,2) band and 
Q$_1$(22)+P$_{12}$(13) of the (7,\,3) band \cite{davis63}.  
Within each of these CN sub-bands there are neighbouring bands which fall
within the 6690--6730\AA \/ wavelength range and should be present
at similar strengths.  These are not seen in the residual spectra,
so it is unlikely that the line detected at 6707\AA \/ is due to CN.  
A similar argument applies to the R$_1$(17)+R$_2$(16)+R$_3$(15) trio 
of lines belonging to the (1,\,4) Swan band
of C$_2$ \cite{phillips68}, which appears at 6707\AA, because the
neighbouring trios at 6703\AA \/ and 6711\AA \/ are not seen.  Other
metal lines are unlikely to produce significant absorption near
6707\AA \/ \cite{barnbaum94}.  Another possible cause of spurious
spectral features is spectral mismatch due to a temperature difference
between the star and the mean spectrum.   A temperature difference of 
say 2000\,K could produce a small apparent absorption at 6707\AA \/ 
due mainly to altering the contribution from the (12,\,7) band,
but it would also gererate larger features at 6722\AA \/ and 6728\AA.  
Since these are not seen, it is reasonable to assume that the 6707\AA \/
feature is not due to a temperature mismatch.  The grouping of the
stars in spectral subclasses helped to minimize any such
problem.  Finally, it is possible that in some cases the presence
of a circumstellar component in the 6707\AA \/ absorption line can
contaminate the photospheric feature.  Although this possibility cannot
be ruled out, it was not found to be a significant effect
in Galactic carbon stars \cite{abia00}.
Therefore, the observed absorption lines at 6707\AA \/
seem to be correctly attributed to lithium.  Ideally, however,
high dispersion spectra of these and some comparison stars would
be needed to confirm the weakest identifications.

\subsection{Stars with detected Li\,{\sc i}\,$\lambda$6707}

%\paragraph{}
Table~1a gives the final sample of carbon stars with
absorption features within 0.75\AA \/ of Li\,{\sc i}\,$\lambda$6707.8, 
with W$_{6707}$\,$>$0.3\AA \/ and with 
2.0\AA\,$\leq$\,{\sc fwhm}\,$\leq$\,5.0\AA.
As seen in the previous section, features with W$_{6707}$\,$>$\,0.4\AA \/ 
are likely to be due to Li\,{\sc i}\,$\lambda$6707 at the 90 per cent 
confidence level, while those with 0.3\,$<$\,W$_{6707}$\,$<$\,0.4\AA \/ 
are likely to be due to Li\,{\sc i}\,$\lambda$6707 at the 50--70 per cent 
confidence level.  Column~1 gives the identification number 
of the star in the Kontizas et al. \shortcite{kontizas01} catalogue, 
column~2 gives the spectral group assigned to each star (see Section~3.1), 
columns 3 to 6 give various photometric measurements which will be 
explained in Section~4, columns 7 and 8 give the W$_{6707}$ (equivalent 
width) and {\sc fwhm} measurements 
of the line at 6707\AA, column~9 gives the j-index taken from 
Morgan et al. \shortcite{morgan03} and column~10 provides cross references 
to Blanco, McCarthy \& Blanco \shortcite{blanco80}:~BMB, Blanco \& 
McCarthy \shortcite{blanco90}:~BM, Hughes \shortcite{hughes89}:~SHV, and
Westerlund et al. \shortcite{westerlund78}:~WORC. 

%\paragraph{}
Table~1b provides a list of LMC and SMC carbon stars with identification of 
Li\,{\sc i}\,$\lambda$6707 recorded in the literature and makes a 
noteworthy comparison with Table~1a. SPLL95 reported four Li-rich LMC carbon 
stars and a further 31 carbon stars for which lithium could not be 
detected.  Richer, Olander \& Westerlund \shortcite{richer79}
and Richer \shortcite{richer81} detected the lithium line 
in two out of 52 stars and two out of 25 stars respectively.  
With one in each of the two pairs common to SPLL95, the total number of
stars is six.  SPLL95 also note two Li-rich SMC stars.  These are both
members of the Rebeirot, Azzopardi \& Westerlund catalogue \cite{rebeirot93}
and both are J~stars as seen in other 2dF data (Cannon et al., 
in preparation).  They are also included in Table~1b.

 \begin{table*}
 \begin{minipage}{140mm}
 \begin{center}
 \caption{Magellanic Cloud carbon stars with Li\,{\sc i}\,$\lambda$6707}
 \begin{tabular}{ccccccccll}
 \hline
 Star & Gp & J & K & $M_{\rm bol}$ & $T_{\rm eff}$ &          $W_{6707}$
  & {\sc fwhm} & j-index & Other \\
  & & & & & & & & & \\
 \multicolumn{5}{l}{a. 2dF stars} \\
1310 & 1 & 11.37 &  9.99 & -5.48 & 3230 & 0.49 & 3.0 &     & \\
1393 & 4 & 12.58 & 10.74 & -4.46 & 2670 & 0.50 & 4.0 & 5.9 & \\
1402 & 1 & 11.79 &  9.86 & -5.32 & 2580 & 0.40 & 3.0 &     & WORC 50 \\
1608 & 4 & 13.54 & 12.46 & -3.34 & 3740 & 0.33 & 2.5 & 3.6 & \\
1792 & 1 & 12.52 & 10.97 & -4.37 & 2990 & 0.60 & 4.0 &     & \\
2155 & 4 & 12.61 & 10.83 & -4.39 & 2730 & 0.48 & 3.5 & 6.0 & \\
2197 & 1 & 11.58 &  9.76 & -5.44 & 2690 & 0.30 & 2.5 &     & BMB BW54 \\
2364 & 4 & 13.04 & 11.20 & -4.00 & 2670 & 0.30 & 3.5 & 5.3 & \\
2400 & 1 & 12.86 & 11.34 & -4.02 & 3030 & 0.45 & 4.0 &     & \\
2406 & 1 & 11.37 &  9.77 & -5.53 & 2930 & 0.35 & 3.0 &     & \\
2832 & 4 & 11.76 & 10.34 & -5.09 & 3170 & 1.42 & 5.0 & 5.6 & SHV 0515089-701859 \\
3344 & 1 & 11.24 &  9.62 & -5.67 & 2910 & 0.31 & 3.0 &     & WORC 138 \\
3495 & 2 & 12.53 & 10.91 & -4.38 & 2910 & 0.44 & 2.5 &     & \\
3525 & 1 & 13.29 & 12.10 & -3.56 & 3530 & 0.30 & 3.0 &     & \\
4404 & 1 & 12.53 & 10.84 & -4.42 & 2830 & 0.35 & 4.5 &     & \\
4141 & 4 & 12.06 & 10.56 & -4.81 & 3060 & 0.49 & 3.5 & 5.4 & \\
4626 & 4 & 12.82 & 10.84 & -4.33 & 2530 & 0.36 & 5.0 & 4.8 & \\
4850 & 2 & 12.54 & 11.19 & -4.31 & 3270 & 0.48 & 3.0 &     & SHV 0534304-720850 \\
5455 & 1 & 12.65 & 11.13 & -4.23 & 3030 & 0.59 & 3.5 &     & BM 35-15 \\
 \multicolumn{5}{l}{b. Published stars} \\
1604 & 4 & 11.80 & 10.36 & -5.00 & 3140 & (4.43) &   &    & WORC 65 \\
2041 & 4 & 13.30 & 11.72 & -3.56 & 2960 &        &   &    & BMB BW9 \\
2809 &   & 12.33 & 10.91 & -4.47 & 3170 & (1.71) &   &    & HV 5680 \\
     &   & 11.98 &  9.65 & -5.60 & 2250 & (6.41) &   &    & BMB R46  SHV 0521050-690415$^\ast$ \\
     & 0 & 11.48 & 10.23 & -5.29 & 3430 & (8.83) &   &    & BMB BW89$^\ast$ \\
     & 2 & 12.10 &  9.35 & -6.22 & 1980 &        &   &    & WORC 186$^\ast$  SHV 0535323-710211 \\
     & 4 & 12.64 & 10.98 & -4.71 & 2817 & (2.45) &   &    & RAW\,1296$^\ast$ \\
     & 4 & 13.02 & 11.29 & -4.37 & 2740 & (1.55) &   &    & RAW\,1329$^\ast$ \\
 \hline
 \multicolumn{10}{l}{$W_{6707}$ and {\sc fwhm} are in \AA} \\
 \multicolumn{10}{l}{$W_{6707}$ values given in brackets were determined 
  by a different method (see text)} \\
 \multicolumn{10}{l}{$\ast$: Gp 0 represents an SC star \cite{richer80}} \\
 \multicolumn{10}{l}{$\ast$: Gp 2 for WORC 186 is assumed from data by 
                     Richer et al. \shortcite{richer79}} \\
 \multicolumn{10}{l}{$\ast$: BMB R46 is also F4488 as recorded by SPLL95} \\
 \multicolumn{10}{l}{$\ast$: RAW\,1296 and RAW\,1329 are SMC members} \\
 \end{tabular}
 \end{center}
 \end{minipage}
 \end{table*}

\subsection{Comparison with other work}

First there is the question of commonality between the 2dF-selected 
and previously known Li-rich stars.   It turns out that not one of 
the six known Li-rich stars were included in the 2dF sample.  Nor were any 
of the stars reported by SPLL95 as having no lithium detection included 
in the 2dF sample.  However, five of the other stars observed by 
Richer et al. \shortcite{richer79} (WORC\,54 $\equiv$ KDM\,1490, 
WORC\,78 $\equiv$ KDM\,1814, WORC\,106 $\equiv$ KDM\,2518, 
WORC\,192 $\equiv$ KDM\,5314, WORC\,196 $\equiv$ KDM\,5426) and three 
of those observed by Richer \shortcite{richer81} 
(BMB\,BW\,31 $\equiv$ KDM\,2117, BMB\,BW\,41 $\equiv$ KDM\,2148, 
BMB\,BW\,54 $\equiv$ KDM\,2197) are in the 2dF sample.  Whereas 
Li\,{\sc i}\,$\lambda$6707 was not noted by these authors as being present 
in any of these eight stars, it was found in one of the 2dF spectra
(KDM\,2197); none of the others has a line anywhere near the limit 
for inclusion in Table~1a.   In fact, the line identified in 
KDM\,2197 is at the lowest limits of both W$_{6707}$ and {\sc fwhm} 
for inclusion in the table.  It is likely that this strength is below 
the limit for identification by Richer \shortcite{richer81}.

%\paragraph{}
The equivalent widths, W$_{6707}$, recorded in Table~1a
range between 0.3\AA \/ and 0.6\AA \/ with one stronger line at 1.4\AA.
Torres-Peimbert and Wallerstein \shortcite{torres66} 
found similar strengths for W$_{6707}$ of $\sim$0.4--0.5\AA \/ in 
fourteen of the sixteen Galactic carbon stars in which they detected lithium.  
The other two were the well known super-Li-rich stars WZ\,Cas and WX\,Cyg.
However, in the more recent studies by SPLL95 in the Magellanic 
Clouds and by Boffin et al. \shortcite{boffin93} in the Galaxy, 
significantly higher values W$_{6707}$ have been reported.  In fact, 
three of the four Li-rich LMC carbon stars described by SPLL95
are in the super-Li-rich category.  Part of the discrepancy between 
levels of W$_{6707}$ is due to the different methods used.  
SPLL95 used a pseudo continuum fitted near 6700\AA \/ and did not attempt 
to correct their values of W$_{6707}$ for a CN contribution.  It is 
worth stating that the W$_{6707}$ levels determined by 
Torres-Peimbert and Wallerstein \shortcite{torres66} and noted above as 
being similar to those of Table~1a were corrected for a contribution
from CN, though through a method different to the one used here.
Inspection of 
Fig.~\ref{montage}c shows that this technique would give a larger 
equivalent width than the one quoted in Table~1a.  Tests on the 2dF 
spectra of J~stars imply differences of 0.4--0.5\AA \/ between the two 
methods.  However, this level of discrepancy is not sufficient to explain 
why no super-Li-rich carbon stars were discovered in the 2dF sample. 

\section{Photometry}
%\setcounter{paragraph}{0}
%\paragraph{}
Infrared photometry of sources in much of the LMC is now available
through the Second Incremental Data Release of the Two Micron All
Sky Survey -- \mbox{(2-MASS)} \cite{skrutskie97} and is available on-line 
at http://irsa.ipac.caltech.edu/.  \mbox{2-MASS} $JHK$ measurements
were extracted for the stars in the 2dF sample by matching catalogue
coordinates.  Whenever \mbox{2-MASS} data were unavailable, $JK$
measurements were taken from the Deep Near-Infrared Survey of the 
southern sky -- DENIS \cite{cioni00} which is available on-line from 
the CDS at Strasbourg (http://cdsweb.u-strasbg.fr/denis.html).  
The magnitudes from the DENIS database were corrected in the way described
by Morgan et al. \shortcite{morgan03} to bring them
into alignment with the \mbox{2-MASS} data.  
%$RI$ photometry of the stars was 
%taken from Kontizas et al. \shortcite{kontizas01}. 

%\paragraph{}
The $J$ and $K$ magnitudes were subsequently used to derive
bolometric magnitudes ($M_{\rm bol}$) and effective temperatures
($T_{\rm eff}$) by applying the relations
$M_{\rm bol} = K_{\rmn o}-dm+0.69+2.65(J-K)_{\rmn o}-0.67(J-K)_{\rmn o}^2$ and 
$T_{\rm eff} = 7070/[(J-K)_{\rmn o}+0.88]$, derived by Wood, Bessell \& Fox
\shortcite{wood83} and  Bessell, Wood \& Lloyd Evans
\shortcite{bessell83}, respectively.  A distance modulus value of
18.45 was adopted \cite{westerlund97} and a mean reddening
of $E(J-K)$\,=\,0.07 \cite{costa96}.  Small deviations from the
adopted value of the reddening do not significantly affect the
derived values of $T_{\rm eff}$ and $M_{\rm bol}$. The resulting values of
$T_{\rm eff}$ and $M_{\rm bol}$ are listed in Columns 5 and 6 of Table~1.
It must be pointed out that the use of these parameterizations for 
$M_{\rm bol}$ and $T_{\rm eff}$ may not be fully applicable to the specific 
stars studied here and must be treated with caution.  Formal errors 
for $M_{\rm bol}$ and $T_{\rm eff}$ are $\pm$0.2\,mag and $\pm$70\,K.   

\begin{figure}
\vspace{8.5cm} 
\includegraphics{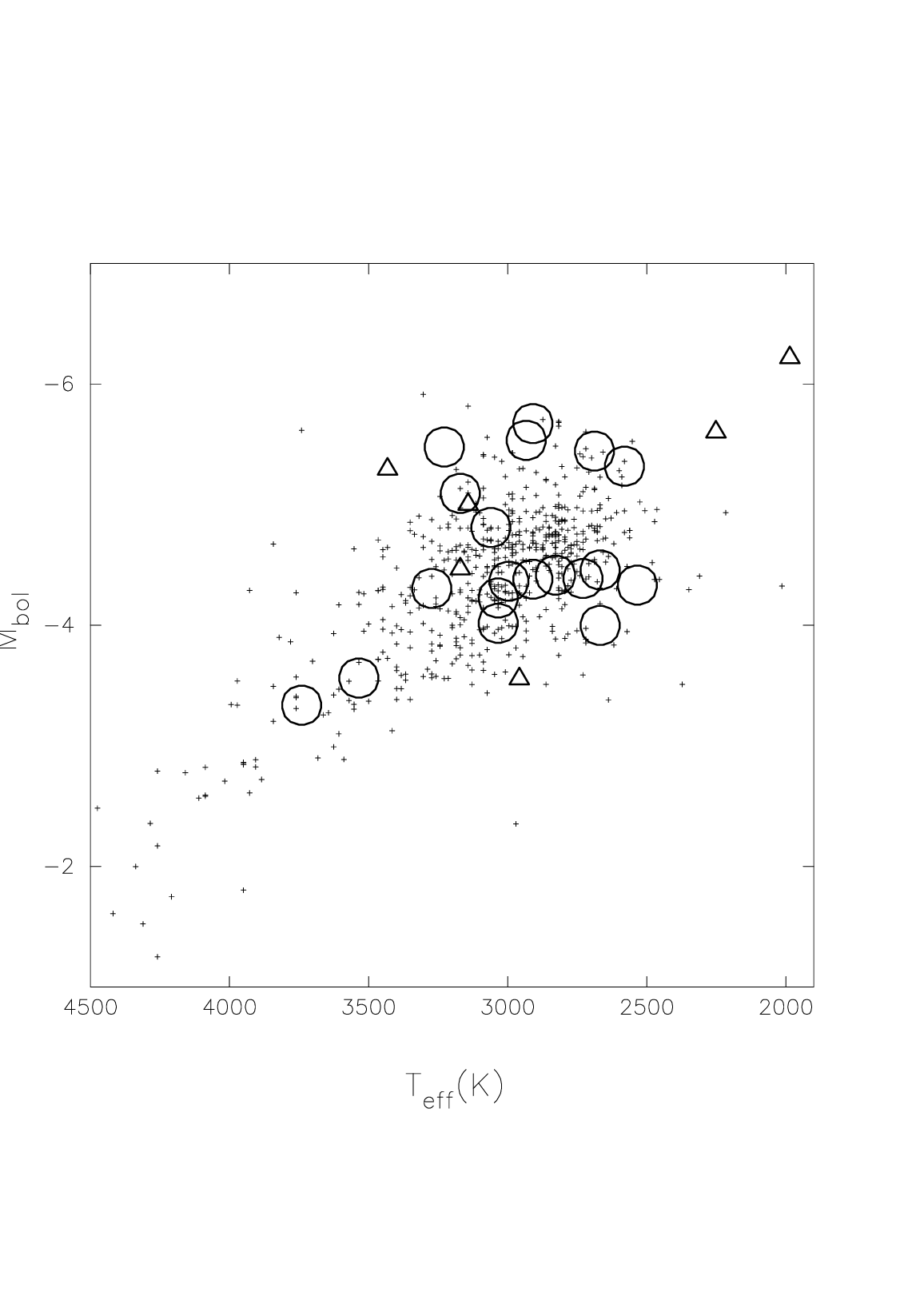}
\caption{A plot of $M_{\rm bol}$ against $T_{\rm eff}$ for LMC carbon stars. 
The circles are the Li-rich stars of Table~1a and the triangles are 
other Magellanic Cloud stars known to have strong
Li\,{\sc i}\,$\lambda$6707 (SPLL95).} 
\label{mbolteff}
\end{figure}

The bolometric magnitudes of the Li-rich stars identified in
the 2dF sample lie in the range $-$3.3 to $-$5.7 which includes fainter
stars than the range of $-$4.6 to $-$5.7 occupied by the four 
Li-rich LMC stars identified by SPLL95.   Similarly faint Li-rich 
carbon stars have been found in the SMC (SPLL95), in M31 
\cite{brewer96}, and in the Galaxy where  
$-$3.5\,$<$\,$M_{\rm bol}$\,$<$\,$-$6.0 \cite{abia97}.

%\paragraph{}
%The mean effective temperature of the Li-rich carbon stars of
%Table~1 is $2967\pm69$\,K.  This is close to the mean value of 
%$2997\pm258$\,K for the four Li-rich carbon stars of SPLL95, but
%somewhat higher than the mean
%value of $2798\pm66$\,K found by Abia \& Isern \shortcite{abia97}
%for 13 Galactic Li-rich stars.  It is not clear if this difference
%is significant, as Abia \& Isern used a different formulation
%(that of Heng et al. (ref?) to derive $T_{\rm eff}$ from 
%the $(J-K)$ colour index.

\begin{figure}
\vspace{6.0cm} 
\includegraphics{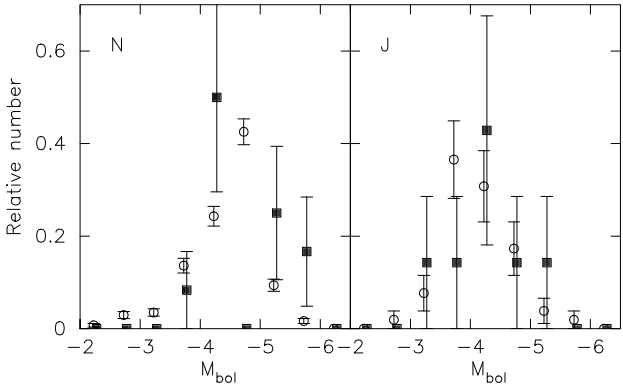}
\caption{Relative frequency of carbon stars in $M_{\rm bol}$ intervals of
0.5\,mag.  The left-hand panel is for N stars and the right-hand panel is 
for J stars.  The open circles are for normal carbon stars and the 
solid squares are for the Li-rich stars in Table~1a.}
\label{histmbol}
\end{figure} 

%\paragraph{}
Figure ~\ref{mbolteff} shows the location of the Li-rich carbon stars 
of Table~1a on
the Hertzsprung-Russell Diagram. For comparison the location of
all carbon stars observed are shown on the same diagram. Their
bolometric magnitudes and effective temperatures were derived in
the same way as those of the stars in Table~1, based on the data in
Cannon et al. (2003: in preparation). The six
Li-rich stars listed in Table~1b are also included in this plot. 
The distribution of the Li-rich stars in Fig.~\ref{mbolteff} is clearly
not the same as that of the normal carbon stars.  Of the stars plotted
as triangles, the two `coolest' (WORC\,186 and BMB R46) are both thought 
to be Mira variables and have the appropriate extreme colours, and the
`warmest' (BMB BW89) is actually an SC star \cite{richer80} and perhaps 
should not be included.  Two of the 2dF Li-rich stars are fainter and 
cooler than the others and lie near the tip of the red giant branch rather
than on the AGB; the rest lie among the main body of AGB stars but seem 
to be more widely distributed and relatively uncommon at the peak of the
general distribution.  Fig.~\ref{histmbol} shows the distribution of 
Li-rich stars with bolometric magnitude (in 0.5-magnitude bins), 
separately for N and J types.  For 
comparison, the distributions for normal carbon stars are superimposed.
The normal N and J~stars stars have clear single peaks at 
$M_{\rm bol}$\,=\,$-$4.75  and $M_{\rm bol}$\,=\,$-$3.75 respectively.  
Although the total number of J stars is small here, studies of a much larger 
sample show that the J stars are indeed fainter than the N~stars 
\cite{morgan03}.   The error bars shown in Fig.~\ref{histmbol} are based 
on $\sqrt n$ statistical errors.  There would not seem to be a 
significant difference 
between the distributions of the bolometric magnitudes of the Li-rich
and normal J~stars.  However, the Li-rich N~stars are absent from the
$M_{\rm bol}$\,=\,$-$4.75 bin which is the peak of the normal N-star 
distribution.  There are, in fact, no Li-rich N~stars in the range
$-$5.25\,$<$\,$M_{\rm bol}$\,$<$\,$-$4.5.

%\begin{figure}
%\vspace{10.0cm}
%\special{psfile=mbol.eps hoffset=80 voffset=-70 vscale=50 hscale=50}
%\caption{Proportion of Li-rich stars among carbon stars in 0.25-mag 
%bins of $M_{\rm bol}$. }
%\label{W_Mbol2}
%\end{figure}

\section{Discussion}

\subsection{Statistics}

%\setcounter{paragraph}{0}
%\paragraph{}
Of the 667 stars studied here, 53 were excluded as being of insufficient
quality to allow the reliable non-detection (or detection) of 
Li\,{\sc i}\,$\lambda$6707 to a limit of W$_{6707}$\,$\simeq$\,0.3\AA.  
Of the remaining 614 stars, nineteen (i.e. 3.1 per cent) have detectable
Li\,{\sc i}\,$\lambda$6707.  Among the 614 stars there are 62 J~stars, 
of which seven are Li-rich.   This means that, to
the detection limit of the present work, 11.3~per~cent of the J~stars
are Li-rich compared with only 2.2~per~cent of the non-J~stars.  So lithium
enrichment seems to be about five times more frequent among J-type
carbon stars. No super-Li-rich carbon stars were discovered
in the sample of 614 stars.  It is probably true to say that none was 
discovered in the full sample of 667 stars because most of the extra 
53 stars could have revealed a very strong lithium line with 
W$_{6707}$\,$\sim$\,4\AA.  

These percentages are significantly lower than those recorded in 
the literature for the Galaxy and the Magellanic Clouds, although 
in the latter case in particular the samples are one or two orders of 
magnitude smaller than the present one.
Boffin et al. \shortcite{boffin93} found that in a sample of about
200 Galactic carbon stars $\simeq$12.5~per~cent were Li-rich and 
1.5~per~cent were super-Li-rich.  (The Boffin et al. \shortcite{boffin93} 
results are updates of generally similar results based on smaller samples, 
presented in the earlier papers of Abia et al. (1991, 1993).)  
More recently, Abia \& Isern \shortcite{abia97} made an extensive 
study of 44 Galactic carbon stars (28 of N type, 11 of J type, and 
5 SC stars) and detected $^7$Li in 82~per~cent of the J~stars
and in 15~per~cent of the rest.  In the Magellanic Clouds, SPLL95 found 
four Li-rich carbon stars out of a total of 35 LMC carbon stars in 
their sample (11.4~per~cent) and 2 out of 25 (8~per~cent) for the SMC.

Part of the difference in the frequencies of Li-rich carbon
stars reported by the different authors may be related to actual
differences between the carbon star populations in the parent
galaxies (ages, masses, metallicities), but part of it may be due
to the different approaches adopted for the lithium detection in
the spectra, and, most importantly, the correction (or lack of it) for 
the CN contribution and the choice of continuum.   The size of the sample 
considered in each case and its initial selection must also be taken 
into account. 

%It has
%already been mentioned in the previous section that the equivalent
%widths of the lithium lines found in the present paper have been
%corrected for CN contributions and are generally similar to the
%CN-corrected equivalent widths of Torres-Peimbert \& Wallerstein
%\shortcite{torres66}, but lower than the Boffin et al. 
%\shortcite{boffin93}, Abia et al. (1991, 1993)
%and SPLL95 equivalent widths which however were not corrected for CN. 
%Actually, Boffin et al. \shortcite{boffin93}
%mention that their equivalent widths are in good agreement with
%those of Torres-Peimbert \& Wallerstein \shortcite{torres66}.  However, 
%they used the uncorrected values of the equivalent widths given by these
%authors for the comparison. In any case, it is obvious that if we
%had not corrected for the CN contribution, the equivalent widths
%would be larger and a lot more stars would pass the border line
%for lithium detection, therefore the frequencies derived would be
%erroneously higher.  We conclude that in the LMC, lithium enrichment in
%the atmospheres of carbon stars is probably less common than
%previously thought, and certainly less extreme than for Galactic
%carbon stars in the sense that no super-LI-rich carbon stars were found
%in the 667 stars studied here. [IS THIS SECOND-LAST SENTENCE CORRECT?]    

\begin{figure}
\vspace{6.0cm} 
\includegraphics{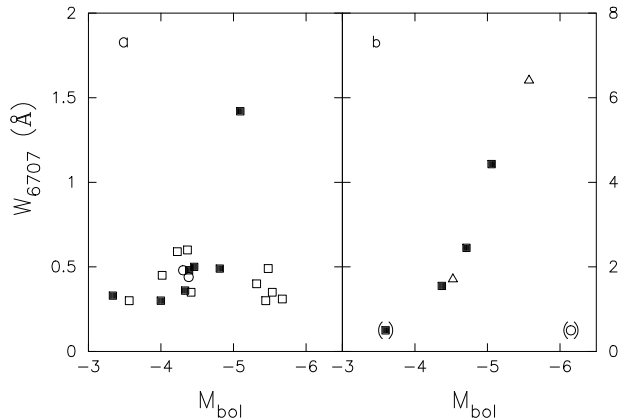} 
\caption{$W_{6707}$ against $M_{\rm bol}$.  Panel (a) is for the stars of 
Table 1a (2dF results) and Panel (b) is for the stars of Table 1b 
(SPLL95 data).  The symbols are: J stars -- solid squares; N stars
-- open squares; C$_2$-weak stars -- open circles; unknown types --
open triangles. Note the different ordinate scales.} 
\label{W_Mbol}
\end{figure}

%\begin{figure}
%\vspace{7.5cm} 
%\special{psfile=wteff.eps hoffset=0 voffset=-90 vscale=45 hscale=45} 
%\caption{$W_{6707}$ against $T_{\rm eff}$ on a logarithmic scale. The
%symbols are as shown in the inset, with the solid symbols representing
%the stars identified in the present work and the open symbols representing
%those from SPLL95.} 
%\label{W_Teff}
%\end{figure}

\subsection{Spectral properties}

%\paragraph {}
In order to gain some insight into the formation of the observed
population of Li-enriched carbon stars, it is useful to investigate the
possible existence of correlations between the equivalent widths
W$_{6707}$ and other properties of the stars, such as bolometric
magnitude, effective temperature, and spectral group (Groups 1--4 as
shown in Table~1a).  Figs~\ref{W_Mbol}a and \ref{W_Mbol}b show
W$_{6707}$ plotted against $M_{\rm bol}$ for the 2dF results in Table~1a and
the data from SPLL95 (Table~1b) respectively.  The SC star, BMB\,89, is
considered to be too different in spectral type and is not shown.
The two points in brackets are estimates of W$_{6707}$ based
on the Li-indices published by Richer et al. \shortcite{richer79}
and Richer \shortcite{richer81}.  No statistically significant 
correlations can be found for the N~stars in Fig.~\ref{W_Mbol}a,
but the bimodality of the distribution can be seen again.  Nor did 
Abia et al. \shortcite{abia93} find any correlation between bolometric 
magnitude and lithium abundance for their sample of Galactic carbon 
stars.  However, if only J-type carbon stars are taken into account, 
a different picture seems to emerge with W$_{6707}$ appearing to increase 
with bolometric magnitude.  For the 2dF sample plotted in 
Fig.~\ref{W_Mbol}a, this conclusion is based on just one star 
and cannot be taken as more than tentative; but it is more striking for the
SPLL95 sample plotted in Fig.~\ref{W_Mbol}b.  As noted earlier, the 
apparent offset between the new data and the SPLL95 results can be 
partly explained by the different continua adopted.
In interpreting these diagrams it should 
be remembered that some of the stars, especially those in the SPLL95 
sample, are variables and could change in $M_{\rm bol}$ and 
$T_{\rm eff}$; e.g.,
BMB~R46 appears in Table~1b with $M_{\rm bol}$\,=\,$-$5.6 and is recorded 
at $M_{\rm bol}$\,=\,$-$6.3 by Whitelock \& Feast \shortcite{whitelock00}.
With these considerations in mind, it is clear that a
larger statistical sample is needed to confirm this result.

%\paragraph {}
So it would appear that J-type stars hold some special position in
the Li-enriched carbon star sample: (i) it seems to be about 5
times more likely to get a Li-enriched carbon star if it is also
of J-type and (ii) there seems to be a correlation between
bolometric magnitude and equivalent width of the lithium line for these
stars.  It must also be emphasized that the highest equivalent width 
of the lithium line in the 2dF  sample occurs in a J-type carbon star.

W$_{6707}$ does not seem to display any obvious correlation with either
$T_{\rm eff}$ or, for the J~stars, with the j-index derived by Morgan et al. 
\shortcite{morgan03} and listed in Table~1a.  However, it must be 
remembered that the size of the sample of J-type Li-enriched stars is 
just seven.

\subsection{Theoretical models}

%\paragraph {}
Given that for the first time the search for Li-enriched carbon
stars has been conducted within such a large and homogeneous
spectral sample, it is worth discussing the theoretical
implications of the results presented here. Ventura et al.
\shortcite{ventura99} produced a self-consistent description of
time-dependent mixing, overshooting and nuclear burning in AGB
stars, and showed that there is a narrow range of stellar masses
in which, despite the activation of the Cameron-Fowler mechanism,
carbon can survive long enough in the convective envelope to keep 
the C/O ratio over unity as well as increasing
$^{13}$C/$^{12}$C, and that consequently these objects can 
be observed as Li-rich C stars for a non-negligible fraction of their
AGB life.  Their model can account for:\\
(i) the very rare luminous Li-rich carbon star with $M_{\rm bol}=-6.5$. No
such star exists in either our or SPLL95's sample of LMC
Li-rich carbon stars. \\
(ii) Li-rich J-type carbon stars with $M_{\rm bol}$\,$\simeq$\,$-$5.75 to
$-5.95$.   The brightest J-type Li-rich star in our sample has 
$M_{\rm bol}=-5.1$, which is more than half a magnitude fainter than required,
for its formation to be attributable to this mechanism.\\
(iii) Li-rich non-J carbon stars for short periods during
thermal pulses. Although this scenario could explain some of the
N-type Li-rich carbon stars, it cannot explain all of them, as the
mechanism can operate successfully only for short periods. Not
enough details are given by  Ventura et al. \shortcite{ventura99} 
to allow for a stronger statement. In any case, the faint J-type 
Li-rich carbon stars in our sample cannot be explained in this way. To
summarize, it would seem that only very few, if any, of the Li-rich
carbon stars of Table~1a can be accounted for by these models; the 
predicted overall distribution of Li-enriched stars is not a good 
match to the observations.  It should be noted that the models which 
reproduce the characteristics of J-type stars suffer from another 
deficiency; they predict that these stars should show enhancements of
the s-process elements because they are thermally pulsing AGB 
stars which suffer the third dredge-up, but no s-process 
enhancements are seen in J-type carbon stars in the Galaxy. Thus, there 
is a strong possibility that J-type stars are not in the AGB phase 
and their chemical peculiarities are due to a different atmospheric
processes.

%\paragraph {}
Travaglio et al. \shortcite{travaglio01} investigated the behaviour
of AGB models -- again in the framework of the HBB process -- with
masses in the range 4--6M$_{\odot}$ and metal abundances between
Z\,=\,0.004 and Z\,=\,0.02. They showed that the condition C/O\,$>$\,1
together with high $^7$Li (log$\epsilon(Li)\simeq4$) can only be
achieved for the 4M$_{\odot}$ Z$=$0.004 case, while much lower Li
abundances (log$\epsilon(Li)\simeq2$) are found for the same mass,
but for Z$\,=\,$0.008.  However, their models are far too bright
($-7<M_{\rm bol}<-6$), to be applicable for the cases of Table~1a. They
do, however, mention that preliminary calculations showed that lithium
enrichment of AGB atmospheres through the HBB process can operate
at lower masses (and give lower $M_{\rm bol}$) for low metallicities
(Z\,=\,0.0001). This statement holds promise with regard to the faint
Li-rich carbon stars in the Magellanic Clouds, but one has to
await for more detailed models.

%\paragraph {}
Abia \& Isern \shortcite{abia97} found observational evidence that 
supports the existence of a deep mixing mechanism operating in low-mass AGB
stars that could be responsible for the creation of Li-rich,
J-type, low-mass (and low luminosity) carbon stars.  This mechanism
could account for the fainter J-type, Li-rich stars of Table~1a, but
cannot explain the equally faint N-type, Li-rich carbon stars
found.

\section{Summary}
%\paragraph {}
The results of the present study can be summarized as follows:\\
(1) The incidence of lithium enrichment among carbon stars
in the LMC is much rarer than in the Galaxy, and about five times
more frequent among J-type than among N-type carbon stars: of the 
614 stars studied here, 19 (3.1~per~cent) have detectable 
Li\,{\sc i}\,$\lambda$6707.  Among the 614 stars there are 62 J-type 
stars, 7 (11.3~per~cent) of which are among the Li-rich stars found 
here.  No super-Li-rich carbon stars were found. \\
(2) The equivalent widths of the Li\,{\sc i}\,$\lambda$6707 line were 
measured from spectra corrected in a statistical fashion for the 
contribution from (mostly) CN. 
The values derived are similar to the corrected equivalent widths for 
the lithium line in Galactic carbon stars derived by 
Torres-Peimbert \& Wallerstein \shortcite{torres66} but significantly 
lower than the values derived by SPLL95 in the Magellanic Clouds (whose 
values were not corrected for CN contamination).\\
(3) The bolometric magnitudes of the Li-rich carbon stars range between
$-$3.3 and $-$5.7.  The distribution in $M_{\rm bol}$ of the Li-rich 
N-type carbon stars seems to have two distinct peaks at $\sim-4.25$
and $\sim-5.25$\,mag.\\
(4) J~stars hold some special position in the Li-enriched carbon star 
sample, as they are about five times more likely to be Li-enriched. 
Also, there seems to be some correlation between bolometric magnitude and 
equivalent width of the lithium line for these stars. \\
(5) Existing models of lithium enrichment via the hot bottom burning
process fail to account for the observed properties of the
Li-enriched stars studied here.  A different deep mixing mechanism 
operating in low-mass AGB stars seems to be required, especially for
the faint N~stars that do show Li-enrichment.

\section{Acknowledgements}

The authors are grateful to the staff of the Anglo-Australian Observatory
for assistance with the observations, to the Australian Time Assignment
Committee for the allocation of telescope time, and to the 2-MASS
project team for making the infrared photometry available.

\label{lastpage}

\end{document}